\documentclass[preprint2]{aastex}

\shorttitle{Velocity Variability in GGSS Stars}
\shortauthors{Bizyaev et al.}

\begin{document}

\title{The Space Interferometry Mission Astrometric Grid Giant-Star Survey. I.
Stellar Parameters and Radial Velocity Variability}

\author{Dmitry Bizyaev\altaffilmark{1,2}, Verne V. Smith\altaffilmark{1,3},
Jose Arenas\altaffilmark{4}, Doug Geisler\altaffilmark{4}, 
Steven R. Majewski\altaffilmark{5},
Richard J. Patterson\altaffilmark{5},
Katia Cunha\altaffilmark{1,6},
Cecilia Del Pardo\altaffilmark{7},
Nicholas B. Suntzeff\altaffilmark{8}, Wolfgang Gieren\altaffilmark{4} }
\email{dmbiz@noao.edu, vsmith@noao.edu, dgeisler@astro-udec.cl, srm4n@virginia.edu, rjp0i@virginia.edu, katia@on.br, cdelpardo@utep.edu, nsuntzeff@noao.edu, wgieren@coma.cfm.udec.cl}

\altaffiltext{1}{National Optical Astronomy Observatory, Tucson, AZ, 85719}
\altaffiltext{2}{Sternberg Astronomical Institute, Moscow, 119899, Russia}
\altaffiltext{3}{McDonald Observatory, University of Texas, Austin, TX 78712}
\altaffiltext{4}{Universidad de Concepcion, Concepcion, Chile}
\altaffiltext{5}{University of Virginia, Dept. of Astronomy, Charlottesville, VA 22903-0818}
\altaffiltext{6}{Observatorio Nacional, Rio de Janeiro, Brazil}
\altaffiltext{7}{University of Texas at El Paso, Dept. of Physics, El Paso, TX 79968} 
\altaffiltext{8}{Cerro Tololo InterAmerican Observatory, La Serena, Chile}

\begin{abstract}

We present results from a campaign of multiple epoch echelle
spectroscopy of relatively faint ($V = 9.5-13.5$ mag) red giants
observed as potential astrometric grid stars for the Space
Interferometry Mission (SIM PlanetQuest). Data are analyzed for 775
stars selected from the Grid Giant Star Survey spanning a wide range
of effective temperatures ($T_{eff}$), gravities and metallicities.
The spectra are used to determine these stellar parameters and to
monitor radial velocity (RV) variability at the 100 m s$^{-1}$ level. 
The degree of RV variation measured for 489 stars observed two or
more times is explored as a function of the inferred stellar
parameters. The percentage of radial velocity unstable stars is found
to be very high -- about 2/3 of our sample. It is found that the
fraction of RV-stable red giants (at the 100 m s$^{-1}$ level) is
higher among stars with $T_{eff} \sim 4500$ K, corresponding to the
calibration-independent range of infrared colors $0.59 < (J-K_s)_0 <
0.73$. A higher percentage of RV-stable stars is found if the
additional constraints of surface gravity and metallicity ranges
$2.3< \log~g < 3.2$ and -0.5 $<$ [Fe/H] $<$ -0.1, respectively, are
applied. Selection of stars based on only photometric values of
effective temperature ($4300\,K < T_{eff} < 4700\, K$) is a simple
and effective way to increase the fraction of RV-stable stars. The
optimal selection of RV-stable stars, especially in the case when the
Washington photometry is unavailable, can rely effectively on 2MASS
colors constraint $0.59 < (J-K_s)_0 < 0.73$. These results have
important ramifications for the use of giant stars as astrometric
references for the SIM PlanetQuest.

\end{abstract}


\section{Introduction}

The Grid Giant Star Survey \citep[GGSS;][]{patterson01} is a
partially-filled, all-sky survey to identify giant stars 
from $9 \lesssim V \lesssim 17.5$ mag 
using the Washington $M, T_2 + DDO51$ filter photometric
prescription outlined in \citet{M00}.  A primary motivation for
the survey and a driver of its design is the selection of stars suitable for
the Astrometric Grid of the Space Interferometry Mission (SIM PlanetQuest). 
These Grid stars, which serve as astrometric references against which the
motions of SIM targets are measured, must themselves be as astrometrically
stable as possible. Thus, they must be free of stellar or significant planetary
companions as well as atmospheric activity (spotting/flaring) that will
induce photocenter wobbles at the several microarcsecond level on of order a
decade-long timescale (approximately the duration of SIM).  G and K giants
were selected by the SIM project as the primary stellar constituent of the
SIM Astrometric Grid because these stars are the most luminous, common
stellar type found in all directions of the sky.  High intrinsic luminosity
places giant stars at greater distances for a given apparent magnitude, and
this increased distance decreases the angular scale of any fixed linear
astrometric wobble, thus making these particular stars less likely to be
problematical as references. The GGSS project has been undertaken with the
specific aim of identifying {\it subsolar} metallicity giant stars, which
are intrinsically more luminous than solar metallicity giants, and also less
likely to have planets \citep[see][and references therein]{fischer05}.

Despite consideration of such effects for careful selection, 
all SIM Astrometric Grid candidates require pre-mission monitoring to assess their
likelihood of astrometric stability during the mission.  
One method is to search for variability in other properties,
such as brightness and radial velocity (RV).  Indeed, we are presently involved, along with
other groups, in a large campaign of echelle high-resolution RV monitoring of Astrometric
Grid candidates. 
Fortunately, not all variability in RV (or luminosity) necessarily translates 
to detrimental astrometric variability. However, some intrinsic 
photometric and/or RV variability is due to processes that, although benign 
in terms of photocenter wobble (e.g., radial pulsations), can mask otherwise 
detectable signatures of problematical sources of variability, like 
astrometric wobbles 
due to planetary, brown dwarf or stellar companions. 
Therefore, because campaigns to monitor radial velocity variability to the 
precision needed for vetting SIM targets (i.e. $\sim100$ m s$^{-1}$) are both expensive and 
time consuming, it is useful to understand how intrinsic RV variability of giant stars
depends on other stellar properties (like temperature, metallicity, and surface gravity)
 to optimize efficient selection of giant stars for the Astrometric Grid.

Although it has long been known that stars at the top of the RGB are both 
brightness and RV-variable \citep[see e.g.][]{pryor88,cote96}, 
there has been little 
systematic study of this variability as a function of stellar atmospheric 
parameters, especially at fainter absolute magnitudes.  
Monitoring of radial velocities for bright red giants
has indicated a high probability of their RV variations
at the level of 100 
m s$^{-1}$ \citep[see][and references therein]{hatzes94}. \citet{jorissen97} 
found that 
red giants with spectral types late G to early K are stable and giants with
later spectral types are all variable.  At the same time, study of Hipparcos
red clump giants \citep{adelman01} reveals photometric stability in this 
evolutionary phase. Radial velocity jitter at large amplitude is expected
for metal-poor, luminous red giants ($M_V \le$ -1.4) \citep[see][]{carney03}.

In their assessment of the suitability of giant stars as SIM Astrometric
Grid members, \citet{frink01} studied a proxy sample of 86 nearby giant 
stars selected from the Hipparcos catalogue.  Their study included three to 
eight epochs of high precision (5-8 m s$^{-1}$) echelle RV measurements on 
timescales from days to a year. \citeauthor{frink01} found that 
most (73 of 84) of the giant stars they investigated
have a stable RV at the level of $<$ 100 m s$^{-1}$, and the peak of the 
distribution of measured dispersions in velocity occurs at 20 m s$^{-1}$.  
However, \citeauthor{frink01} do report that RV variation is more 
probable for redder stars ($(B-V) >$ 1.2 in their sample). While the closest 
match to the survey presented here, the \citeauthor{frink01} survey contains 
stars that are much brighter than the expected SIM Grid sample, and, moreover, 
they did not probe variability as a function of metallicity, which is a key 
aspect of the stars being found in the GGSS.

Our goal here is to carry out an initial RV-stability assessment of stars
more like those expected to fill the Astrometric Grid by focusing on stars
taken directly from the GGSS itself. The hope is that RV stability may 
correlate with some intrinsic stellar property, such as effective temperature, 
surface gravity, or metallicity. Although the precision of our RV
measurements is an order of magnitude less than that of \citet{frink01}, it
is appropriate for finding RV wobbles at the level needed to identify
astrometrically-detrimental grid candidates for the fainter, $>1$ kpc
distant GGSS candidates (and, indeed, is the precision at which monitoring
campaigns of SIM Grid stars are being conducted).
Moreover, our sample is almost an order of magnitude larger than that in 
the \citeauthor{frink01} survey. The observations studied and discussed 
here were obtained under an initial JPL-sponsored program in which the 
internal RV accuracy was set at about 100 m s$^{-1}$.

\section{Observations and Data Reduction}

\subsection{Sample Selection and Biases}

The GGSS finds giant stars by photometry in the Washington $M,T_2+DDO51$
filters according to the methods described in \citet{M00}.  
The $(M-T_2,M-DDO51)$ two-color diagram (2CD) effectively distinguishes 
between
late type dwarfs and giants based on the 
surface-gravity-sensitive Mgb+MgH feature near 5150 \AA\ (where the
130\AA\ --wide DDO51 filter is centered). Photometry was obtained in 1302
evenly placed fields across the celestial sphere, each of area 0.5--1.0
deg$^2$.  The 2CD (i.e., the Mgb+MgH feature) is secondarily sensitive to
metallicity and the positions of stars in the 2CD can therefore be used to
derive a crude estimate of the metallicity of the likely giant stars. 
Metal-poor giants are the most widely separated stars in the 2CD from the
locus of dwarf stars, and therefore these are the easiest to identify with
our methods.  Additional details regarding the GGSS can be found in
\citet{patterson01}.

The selection of stars for the present study is biased according to the 
same specific set of criteria used in the original agreement of the GGSS 
collaboration with the SIM Project 
regarding the selection of Astrometric Grid candidates from the 
GGSS.\footnote{In the 
past year, the SIM Project has modified somewhat the 
criteria for the selection of stars for the Astrometric Grid, with a 
focus on stars with magnitudes more typically $V \sim 9-11$ mag and chosen
from both the GGSS and the Tycho-2 catalogues in a ratio of approximately 1:3. 
The Tycho stars are selected without foreknowledge of metallicity, and so 
are expected to be typically near solar metallicity and at smaller distances than stars 
from the GGSS.  In this respect,
all of our results are still relevant, however, since the stars included in the present study
span the range of properties expected for both GGSS and Tycho-2 stars.}
Giant star candidates
are identified from the 2CD and assigned photometric metallicities therefrom.  From the apparent magnitude, color and metallicity, an estimate of the absolute magnitude and a photometric
parallax distance is assigned \citep{rhee}.  For each field, all giant candidates
with $V<13.5$ mag are rank-ordered according to distance, with the most distant star assigned
highest priority.  The top four ranked stars in each field, yielding more 
than 4000
candidates of $9.5 \lesssim V \lesssim 13.5$ mag, were passed onto both
medium and high resolution spectroscopic observing campaigns (the 
latter forming the database explored here).  Typically, though
not exclusively, the most distant giant candidates in each field were also among those
with the lowest photometric metallicities brighter than $V=13.5$ mag.  Thus, our
spectroscopic sample is biased toward more metal-poor and distant stars than would
be found from a random selection of giant stars in the same magnitude range.  From
the above described sample of more than four thousand stars, 
we explore here a random subsample of 775 stars in both celestial hemispheres,
which constitutes
those stars for which multiple high-resolution spectra have been obtained during the 
first two years of the GGSS follow-up program.

\subsection{Spectroscopy}
\label{spectroscopy}

Spectroscopic observations of 434 stars in the northern subsample of GGSS
candidates were conducted in the period from January 2001 to December 2002.
We made use of the 2.1m telescope at the McDonald Observatory and the
Sandiford Cassegrain echelle spectrograph, which provides $R=55000$
resolution. Th-Ar comparison spectra were taken right before every program star
observation. The quality of the data was monitored by observing one or two
RV standard stars from \citet{nidever02} per night.  The adopted setup of
the spectrograph enabled us to cover the 5000-5900 \AA\, spectral range and
achieve a signal to noise (hereafter $S/N$) level of order 20-40 in
reasonable exposure times (10-30 minutes).

The observed spectra were reduced from two-dimensional to ``echelle" format
in a standard way with the IRAF\footnote{IRAF is distributed by 
NOAO which is  operated by AURA, Inc. under contract with the NSF.} 
software package. 
Corrections for
bias, flat field, and scattered light were applied, and cosmic ray hits 
cleaned out by the IRAF's tasks from {\it crutil} package. 
The internal accuracy of the wavelength calibration via the Th-Ar lamp 
spectra was on average 0.001 \AA\  (and not worse than 0.0018 \AA), which
corresponds to an RV accuracy of order 55 
m s$^{-1}$ (and not worse than 100 m s$^{-1}$).

Resampling the spectra introduces systematic errors in the wavelength 
calibration that degrades the RV accuracy.
Thus, to preserve the wavelength calibration the spectrum orders were
not concatenated to produce a single spectrum, but rather were maintained 
individually at their natural sampling.

Southern hemisphere GGSS stars were observed with the 1.2-m Swiss telescope 
and CORALIE velocimeter in 2001-2004
at the ESO La Silla Observatory (Chile), as described by \citet{arenas02}. 
CORALIE is an improved version of the ELODIE spectrograph \citep{baranne96}. 
The effective resolution of CORALIE is 50000, it covers the wavelength range 
3870 -- 6800 \AA\, and it provides a precision of typically 30 m s$^{-1}$ in 
the radial 
velocities. Typically a $S/N$ of 10 was achieved for program targets. 
The radial velocity and its accuracy comes directly from CORALIE's 
reduction package TACOS \citep{baranne96}.

CORALIE provides RVS for 375 program stars directly from the TACOS software. 
We found that spectra $S/N \le 4$ have systematically low RV accuracy and 
rejected them from further analysis. The whole sample of good $S/N$ CORALIE 
candidates is thus reduced to 341 objects.

While we cannot obtain abundances directly from CORALIE
spectra, we incorporate the RVs of these stars in our analysis and adopt
photometrically estimated stellar parameters for these stars.

Since the southern sample was observed with a smaller aperture telescope 
than the northern one, it is biased toward brighter stars:
the distribution of the southern V-magnitudes peaks at 11.6 mag
whereas the northern sample distribution peak is located at 12.7 mag. 

\section{Data Analysis}
\subsection{Radial Velocities}
\label{RV}

RVs from the McDonald spectra were measured using a cross-correlation 
methodology with the help of IRAF's {\it fxcor} task.
After the primary reduction 
the continuum of each order looks almost flat except at its
edges where the continuum level looks systematically lower and the $S/N$ 
is degraded. 
For the RV analysis we discard the pixels at order edges,
which is about $\sim 20\%$ of the total.

\begin{figure}
\plotone{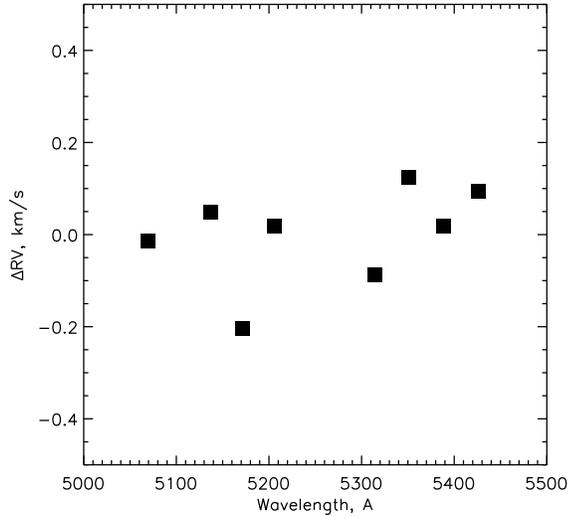}
\caption{ An example of the order-to-order radial velocity differences
for the program star G1113+00.20. The wavelength corresponds to the middle
of each respective order.
\label{fig1}}
\end{figure}

Each spectral order was cross-correlated against the RV template separately.
The template chosen was the
Arcturus spectrum \citep{yellowbible}.  The cross-correlated pieces of 
the template spectrum 
were convolved with a Gaussian corresponding to the FWHM of the Th-Ar
lines found for each order. In addition, the corresponding piece of the 
calibrating Th-Ar spectrum was cross-correlated with a laboratory Th-Ar 
template \citep{th-ar}.
The Th-Ar template for each order was also convolved with the same
Gaussian as for the stellar template in that order.
The shift in the radial velocity between the observed Th-Ar and the laboratory 
reference frame was subtracted from the stellar RV obtained from the 
cross-correlation via the template.
Thus we not only have taken into account the varying
instrumental profile by this order-by-order treatment, but we are 
able to make multiple estimates of the RV for a single observation and 
thereby assess the errors in our velocities for each star.  
A few of the spectral orders indicated
significantly different RVs from the others because of poor wavelength calibration due to 
a sparse Th-Ar line sample. We discarded such "bad" orders from our 
analysis. The remaining orders' radial velocities ($RV_i$) give us 
an independent assessment of the mean RV ($RV_0$) and its accuracy: 
$RV_0 ~=~ \Sigma ~RV_i \cdot w_i ~/~ \Sigma ~w_i $, where $w_i$ is the 
inverse square of the individual RV error provided by IRAF. 
The dispersion of this estimate $\delta RV$ was found as 
$\delta RV ~=~ \sqrt{\Sigma ~w_i \cdot (RV_i-RV_0)^2 ~/~ \Sigma ~w_i} $.
We show an example of the order-to-order radial velocity difference
assessed for the program star G1113+00.20 in Figure \ref{fig1}.

\subsection{Stellar Parameters Derived from an Automated Spectroscopic Analysis}

\begin{figure}
\plotone{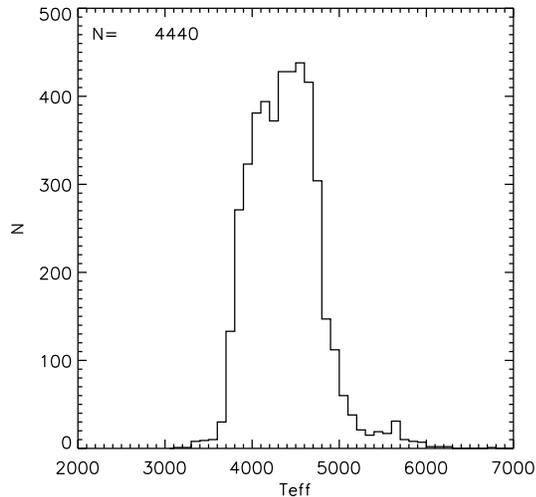}
\caption{ Distribution of $T_{eff}$ for the general sample of 4440 GGSS
candidates. \label{fig2}}
\end{figure}

In addition to the measured RV's, physical stellar parameters can be
derived for the red giants studied here. The effective temperature ($T_{eff}$) of giant stars can be determined
from their broadband colors.  According to an observational calibration
given by \citet{alonso99}, near-infrared colors ($J-K_s$) can be used to 
estimate $T_{eff}$ for red giants, even with no information on the stellar
metallicity and surface gravity.  The full release of the All-Sky Point Source Catalog of
the Two Micron All-Sky Survey (2MASS) makes available the near-infrared 
colors of all GGSS candidates. Figure \ref{fig2} shows the $T_{eff}$ 
distribution of the combined northern plus southern samples as estimated 
from ($J-K_s$) colors according to the Alonso et al. prescription. As may be 
seen, our candidates span a range of temperatures characteristic of red 
giants. The typical accuracy of 2MASS colors is 0.03 mag in our sample 
magnitude range. It corresponds to a 100 K accuracy in $T_{eff}$.

The internal accuracy of the $T_{eff}$ calibration 
in \citet{alonso99} from ($J-K_s$) is of order 125 K. On the other 
hand, \citet{bessell98}
provide a calibration of the $T_{eff}(J-K_s)$ relation for giant
stars from NMARCS stellar atmosphere models. The \citeauthor*{bessell98} 
results are reproduced by the \citeauthor*{alonso99} data fairly well
for typical values of the surface gravity \citep*[see Figure 12 in][]{alonso99}.
The relation from \citet*{bessell98} is metallicity-dependent but the
difference between stars of [Fe/H]=0.0 and -2.0 (which corresponds to the bulk
of our objects, see below) affects $T_{eff}$ by less than 100 K and
would not have a significant effect on the applied calibration procedure.
From the previous two paragraphs, we expect $\sim$ 160 K accuracy 
in our derived $T_{eff}$.

For the northern sample (434 stars) observed at McDonald Observatory, the
reduced echelle spectra can be used to determine the surface 
gravity, $\log{g}$, and metallicity, [Fe/H], by comparison to a library of 
synthetic stellar spectra. For this purpose we produced artificial spectra 
using the MOOG-2002 spectral synthesis package \citep{MOOG}. The 
corresponding stellar models were computed by the ATLAS\,9 code \citep{ATLAS9}.
We computed a model set covering the 3800 -- 5600 K temperature range,
$0.2 \le \log{g} \le 5.6$,
and $-3.0 \le$ [Fe/H] $\le +0.5$
with steps in these parameters of 200 K, 0.2 dex and 0.2 dex respectively. 
In all the models we assumed a typical value for
the microturbulent velocity of 2 km s$^{-1}$.

We selected three pieces of spectra to estimate the stellar parameters.
The Arcturus spectrum (used as an RV template in \S{3.1}) was split 
into 40 \AA\ pieces, and the optimal synthetic spectrum's piece and 
corresponding model parameters were found by least-square fitting.
The stellar parameters of Arcturus (obtained by \citet{griffin99})
were best reproduced in the regions of
5175 -- 5215, 5215 -- 5255, and 5700 -- 5740 \AA .  These wavelength
intervals used to derive the stellar parameters are not simply dominated
by Fe absorption lines, but also by (principally) lines of 
\ion{Mg}{1}, \ion{Ti}{1}, \ion{Cr}{1}, \ion{Ni}{1}, and to a lesser 
degree \ion{V}{1}.  The abundances derived in this
way are therefore not direct ``Fe abundances'', but an overall ``metallicity''
defined by a mixture of both Fe-peak and $\alpha$ elements; we refer to
this abundance as simply metallicity and denote it in the remainder of 
this paper as [Fe/H]. Although significantly non-solar abundance 
ratios, relative to Fe, could create additional noise in the metallicity 
estimates, the vast majority of stars in the sample are well above 
[Fe/H] $\sim$ -2. Within this metallicity regime only Mg/Fe, and to a lesser
extent Ti/Fe and Cr/Fe, are mildly non-solar and typically enhanced. This
may introduce a small trend of slightly larger derived overall metallicities
as the Fe abundance decreases.

To match the wavelength intervals defined above, 
spectra of program stars were corrected to zero radial 
velocity incorporating the RVs defined in \S~\ref{RV}.
The continuum level in the fragments of spectra was fit by a linear 
function using the {\it continuum} task from the IRAF package. The pieces of
synthetic spectra utilized for the comparison were convolved with a Gaussian
corresponding to the FWHM of each echelle order to reduce their spectral 
resolution to that in the observed data. Then we find the best-fit model
for each piece by the least-square method. The values and errors 
of $\log~g$ and [Fe/H] were finally obtained by averaging these same derived 
parameters across the set of three modeled pieces.

\begin{figure}
\plotone{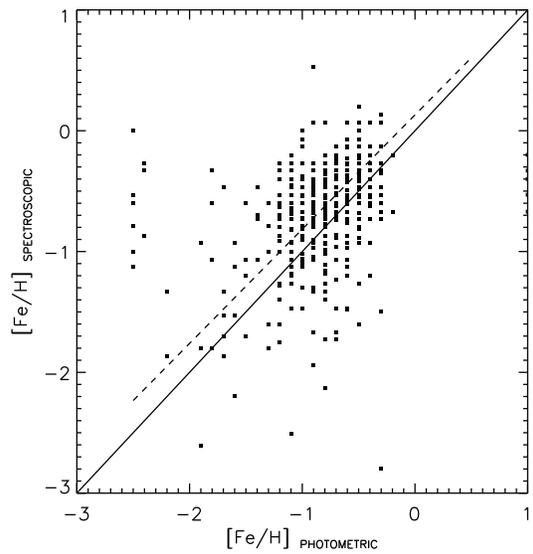}
\caption{ The comparison of metallicities estimated from high-resolution
spectroscopy (see text) and ``photometric" [Fe/H] \citep{rhee}. 
The dashed line shows the ``robust" least absolute deviation fit to the 
data. 
The values of [Fe/H] for most of the stars with intermediate metallicity 
approximately follow a linear relation with dispersion  
of order 0.4 dex. The spectroscopic [Fe/H] is systematically higher
than the photometric one by about 0.16 dex.
\label{fig3}
}
\end{figure}

With this method, we derived the stellar parameters 
($T_{eff}$, $log~g$, [Fe/H]) for the 434 northern GGSS stars (see Table 1). 
The [Fe/H] for the whole GGSS sample has also been estimated 
photometrically \citep{rhee}\footnote{These photometrically derived metallicities are
those that have been delivered by the GGSS collaboration to the SIM project.} 
and one can compare the values derived by each 
method (photometric and spectroscopic).  Figure \ref{fig3} 
shows that the two [Fe/H] estimates for most stars of intermediate 
metallicity ([Fe/H] = -1.0 to  0.0) follows a linear relation with dispersion 
of the order of 0.4 dex, but systematically offset by 0.16 dex (in the sense
that the spectroscopic metallicities tend to be larger). This small 
offset could be due partially to mildly elevated Mg/Fe, Ti/Fe, or Cr/Fe 
ratios in the metal poor stars. There are some extreme outliers in 
Figure \ref{fig3}. Some of these stars have large error bars, either in their
spectroscopic or photometric metallicities. Many of the outlying
points in Figure \ref{fig3} correspond to the hottest stars in our sample. 
These stars occupy
the border region between giants and dwarfs on the giant-dwarf discrimination
diagram from \citet{M00}, and their stellar parameters determined from 
photometric data may be biased.

\begin{figure}
\plotone{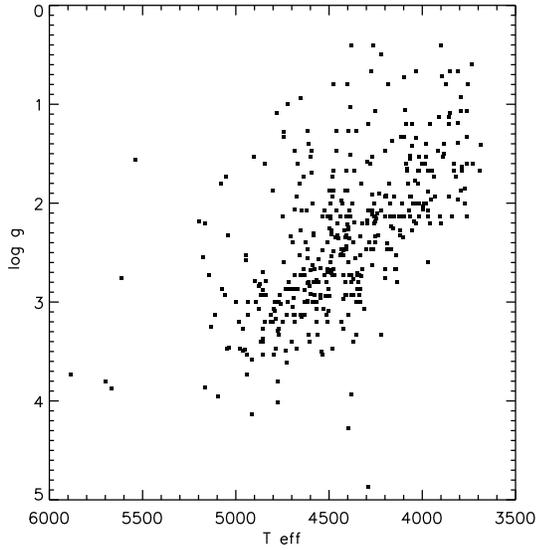}
\caption{ 
The relation between surface gravity and effective temperature for 
434 stars from our northern sample.
\label{fig4}
}
\end{figure}

Figure \ref{fig4} shows the relation between the surface gravity and effective
temperature for the northern hemisphere sample. Most of the stars in 
Figure \ref{fig4} span ranges typical for red giants. Here we see the 
success rate of the GGSS for photometrically identifying bona fide red 
giants to be extremely high: more than 98 \% of the stars that were 
photometrically identified by the \citet{M00} method to be giant stars have 
spectroscopic gravities supporting this characterization.

\subsection{Errors in Atmospheric Parameters}

\begin{figure}
\epsscale{.80}
\plotone{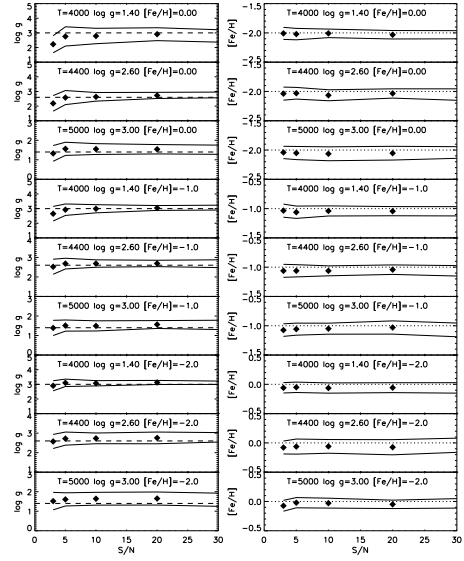}
\caption{ 
The 1-sigma level scatter of estimated stellar atmospheric parameters 
({\it solid curves}) as a function of $S/N$. The mean values of $\log~g$ 
and [Fe/H] found for one hundred deteriorated model spectra
are designated by the diamonds. The dashed lines represents the initial 
(true) parameters ($\log~g$ or [Fe/H]) of the model spectra. The results of 
simulations are shown for models typical of the GGSS red 
giants: $T_{eff}=4000, ~\log~g=1.4$, $T_{eff}=4400, ~\log~g=2.6$, 
and $T_{eff}=5000, ~\log~g=3.0$, and for metallicities [Fe/H]=0, -1, and -2. 
\label{fig5}
}
\end{figure}

In order to analyze the accuracy of the derived stellar parameters as a function of
the quality of observed spectra, we performed Monte Carlo
simulations deriving the basic stellar parameters for a set of selected
synthetic spectra deteriorated by varying amounts of
added noise. We considered a set of signal-to-noise ratios from 3 to 30 which
covers the typical range of $S/N$ in our observed spectra.  
Besides random noise, we also added variations
of $T_{eff}$ and RV into the model spectra, distributing them uniformly
within $\pm$100 K and $\pm$100 m s$^{-1}$ range from the true value. 
Finally, we
assumed that the microturbulence velocity might take an arbitrary value in
the range 1 to 2 km s$^{-1}$. The stellar parameters were defined for a
hundred resulting synthetic spectra over the typical (see Figure \ref{fig4}) 
values of $T_{eff}$ and $log~g$: (4000 K and 1.4 dex), (4400 K and 2.6 dex), 
and (5000 K and 3.0 dex).  
The [Fe/H] took the values -2, -1, and 0 dex for each case. The
resulting ranges of the estimated parameters (1-$sigma$ level) are shown
by the solid curves in Figure \ref{fig5} about the
mean values (designated by the diamonds).
The dashed lines represent the initial true 
parameters ($\log~g$ or [Fe/H]) of the spectra. The figure shows that we 
obtain an accuracy in $\log~g$ of the order of 0.2 dex. A systematic 
difference in $\log~g$ of +0.1 dex
is found and is primarily the result of the displacement of the 
microturbulent velocity $\xi$ from the adopted 2 km s$^{-1}$.
The metallicity is defined with 0.1 dex accuracy and a possible displacement 
of all estimated values (again due to the unknown $\xi$) is -0.05 dex.  To check 
if these results depend on the number of Monte Carlo simulations, 
we also evaluated some models based on sets of 400
deteriorated spectra instead of 100. No obvious
differences can be seen compared with the case of 100 simulations.

We also extend the
considered $S/N$ ratio up to 100. High values of $S/N$ do not improve
the inferred parameters since the $\xi$ uncertainty introduces the most
significant error. The next strongest factor introducing a systematic
deviation between the "real" and estimated stellar parameters is the RV error.
It is seen that starting with $S/N=10$ and better one can estimate 
surface gravity and metallicity from our spectra with quite good
accuracy. Most of our spectra in the northern sample satisfy
this criterion.  Note that the accuracy of
the RVs also depend on $S/N$. This can be a reason for more
significant errors in $\log~g$ and [Fe/H] obtained from our spectra.

\subsection{Stellar Parameters Derived from a Classical Spectroscopic Analysis}

Some of the higher $S/N$ spectra from GGSS stars studied here are suitable 
for a detailed abundance analysis from which stellar parameters and iron 
abundances can be obtained via measurements of individual equivalent widths 
from a selected sample of \ion{Fe}{1} and \ion{Fe}{2} lines. This is a 
different technique 
from the straightforward matching of observed and synthetic spectra used
in Section 3.2, which we will refer to as the ``automated spectroscopic''
method.  When a strictly spectroscopic method (using a line-by-line analysis) 
is adopted, the effective temperatures can be obtained by forcing a zero 
slope in the relation between \ion{Fe}{1} abundances with line  excitation 
potentials; surface gravities are obtained from the agreement between the 
abundances from \ion{Fe}{1} and \ion{Fe}{2} lines. Another parameter that 
can be adjusted at the same time is the microturbulence velocity, which is 
tuned so that the \ion{Fe}{1} abundances are independent of the equivalent 
widths.

This method can provide a consistency check on the metallicities
and $log~g$ derived with the automated method presented in
Section 3.2.  To test the degree of consistency, a small sub-sample 
of 10 target stars, with larger than average $S/N$ values, 
were selected as candidates for the detailed spectroscopic analysis.
Our approach consisted of adopting the same effective temperatures from
the \citet*{alonso99} photometric calibration for the target stars
and to use a sample of \ion{Fe}{1} and \ion{Fe}{2} lines that were tested
to produce good results for the Sun, as well as the well-studied red-giant
Arcturus. The sampled Fe lines and their atomic parameters are listed 
in Table 2: for each transition the excitation potential and gf-value
are listed.  The equivalent-width measurements are presented in Table 3.
The same spectrum analysis code MOOG was used in the abundance computations 
based on the individual equivalent-width measurements. 
The adopted effective temperatures, and derived surface gravities, 
microturbulent velocities, and iron abundances are presented in Table 4.  
A comparison of surface gravities derived using 
the \ion{Fe}{1} and \ion{Fe}{2} lines with those obtained from the automated 
method finds a systematic mean offset 
of [log g(automated) - log g(Fe I/Fe II)]= +0.4$\pm$0.4 dex. 
As the \ion{Fe}{2} lines are quite sensitive to stellar surface 
gravity, this systematic offset may suggest that the surface gravities 
derived from our automated method are somewhat overestimated.
 
\begin{figure}
\epsscale{.75}
\plotone{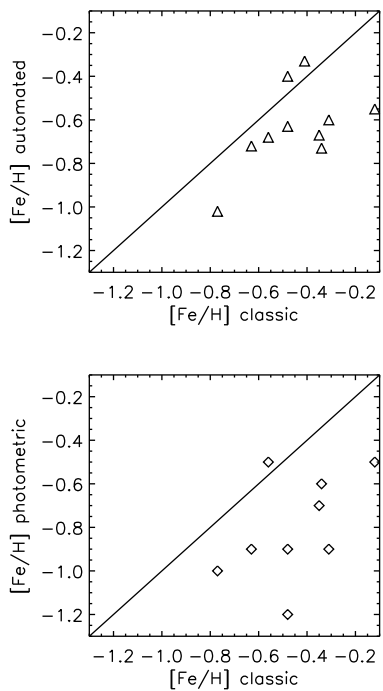}
\caption{
Metallicities for 10 GGSS stars measured three different
ways: using Washington photometry, the automated method, and a detailed 
spectroscopic analysis.  The three methods show general agreement.
\label{fig6}
}
\end{figure}

The derived Fe abundances from this detailed analysis are intercompared in 
Figure \ref{fig6} with the metallicities obtained from Washington + DDO51
photometry (upper panel) and automated spectroscopy (middle panel). The 
metallicity distributions shown in the middle and lower panels of the figure 
are in generally good agreement, with the same standard deviation of 0.18 
dex, but an an offset of 0.18 dex in the average [Fe/H].  Such an offset is
within the quoted uncertainties in the metallicities
obtained from the automated method (of roughly 0.2 to 0.3 dex).  
Moreover, because the automated method does not measure a true Fe abundance, 
but an overall general metallicity affected by a mixture of elements, 
a mild offset between the two methods is not unexpected.

An important aspect of this additional verification of the results obtained 
with the automated method is a check on the derived surface gravities
using Fe I and Fe II lines directly, via ionization equilibrium.
As discussed previously, the GGSS targets were selected from Washington
photometry two-color diagrams to be giant stars and not dwarfs. 
The values of $\log~g$ derived for this subsample of stars, as can be seen in
Table 4, confirms their evolved status as red giants.  Both the direct
analysis of Fe I and Fe II ionization equilibrium, as well as the automated
direct comparison with model spectra, indicate that the Washington + DDO51 
two-color diagram is an effective method for identifying red giants.

\section{RV Stability Versus Stellar Parameters}

We now focus on the 148 stars of the northern sample whose RVs were
measured two or more times with large epoch differences (a time
between successive observations of the same star not less than 4
months). From these data we estimate the RV dispersion which we designate 
as $\sigma$ and utilize this as a measure of variability of the RV. 
Note, this is not the statistical standard deviation widely designated
by $\sigma$; the present $\sigma$ is calculated from multiple RV measures 
after weighting each by the inverse square of the accuracy of each RV 
measure ($w_j$): $RV_m ~=~ \Sigma ~RV_j \cdot w_j ~/~ \Sigma ~w_j $, and
\begin{equation}
\sigma ~=~ \sqrt{\Sigma ~w_j \cdot (RV_j-RV_m)^2 ~/~ \Sigma ~w_j}.
\end{equation}
Here $w_j$ is the weight. If the accuracy of the corresponding RV was
better than 50 m s$^{-1}$, we assume it equals 50 m s$^{-1}$ because of
the wavelength calibration uncertainty (see \S3.2). 

\begin{figure}
\plotone{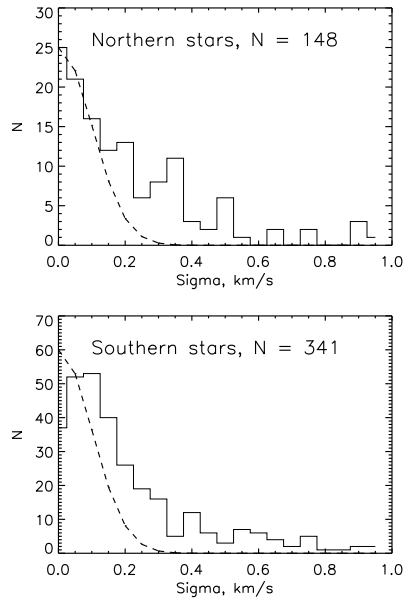}
\caption{ 
Distribution of the variation of radial velocity, $\sigma$ (in km s$^{-1}$) 
for the northern (upper panel)  and southern (lower panel) subsamples (148 
and 341 stars, respectively) with two or more estimates of RV. The dashed line
shows a gaussian distribution with 100 m s$^{-1}$ sigma).
\label{fig7}
}
\end{figure}

The derived values of $\sigma$ are shown in Table 5. 
Figure \ref{fig7} (upper panel) shows the distribution of the $\sigma$ 
values (in km s$^{-1}$). The largest number of sources occur in the first 
bin reveals the accuracy of our estimates and confirms that its value is 
of the order of 50 m s$^{-1}$, as it follows from the accuracy of the
wavelength calibration (section \S2.2).

We next incorporate the repeated RV data obtained with CORALIE 
for the southern GGSS subsample \citep{arenas02} in Figure \ref{fig7} 
(lower panel) with 341 stars in this sample (see Table 6). 
Variation of the radial velocity $\sigma$ for the southern stars observed 
two and more times is estimated by Equation (1).
A gaussian curve corresponding to 100 m s$^{-1}$ sigma is shown
in Figure \ref{fig7} by dashed line.
Both southern and northern subsamples exhibit similar behavior 
in their respective RV-variability distributions.

We found no large, apparent systematic
difference in the stellar parameters between the RV-stable (low $\sigma$)
and unstable (high $\sigma$) fractions of the sample. 
However, for stars with $\sigma < 100$ m s$^{-1}$ the {\it ranges} of the 
stellar parameters appear to be narrower than for red giants 
with $\sigma > 100$ m s$^{-1}$. Thus, the fraction of RV-stable stars 
is low among very cool ($T_{eff}<$ 4000 K) as well as metal-poor stars 
with [Fe/H] $<$ -1.0, in agreement with previous conclusions
by \citet{jorissen97} and \citet{carney03}.

\begin{figure}
\plotone{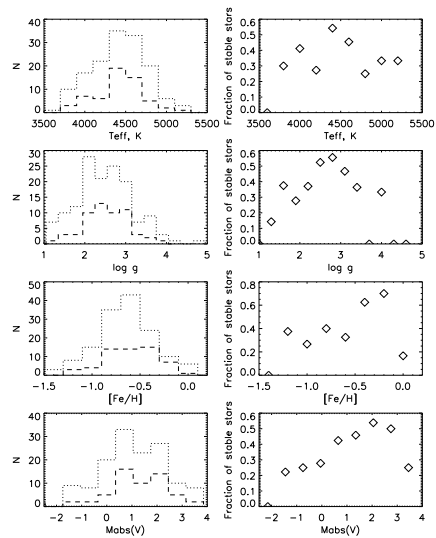}
\caption{ 
Left panels: histograms of distribution of $T_{eff}$, $log~g$, 
[Fe/H], and $M_V$ for the stars whose $\sigma < 100$ m s$^{-1}$ 
(dashed line) in comparison with the whole sample (dotted line). 
The right side panels show the corresponding fraction of stable 
stars in bins of the histogram.
\label{fig8}
}
\end{figure}

The fraction of RV-stable and unstable stars can be estimated from the
distributions of $T_{eff}$, $log~g$, [Fe/H], and absolute magnitude $M_V$
presented in the left-hand panels of Figure \ref{fig8};
here the distributions of stars with $\sigma < 100$ m s$^{-1}$ (dashed 
lines) are shown in comparison with the distributions for all stars 
(dotted lines).  An assessment of $M_V$ is made with $T_{eff}$, $log~g$, and
an assumption of fixed mass for the red giants (we assume it equals to 
0.9 $M_{\odot}$; variation of the mass introduces insignificant scatter 
in $M_V$). The right-hand panels of Figure \ref{fig8} show the fraction of 
stable stars in each bin of the histograms shown on the left.
As seen in the figure, the fraction of stable stars has a peak in the range
4300-4700 K (secondary peaks at the wings of the distribution are caused 
by small number statistics). 
Note that this range corresponds to the extinction-corrected $(J-K)_0$
values from 0.59 to 0.73 mag, and these values do not depend on the choice of
color-temperature calibration. Figure \ref{fig2} shows that the bulk of the 
giant stars selected from the GGSS Washington$+DDO51$ photometry survey 
occupies this range of temperatures, showing that the GGSS happens to be 
optimized for selecting stars in the most RV-stable temperature range.

A similar peak takes place at $\log~g$ = 2.3 -- 3.2 in the surface gravity
distribution. We also find that the largest fraction of RV-stable stars 
happens for stars of slightly subsolar metallicity (-0.5 $\le$ 
[Fe/H] $\le$ -0.1), and then the fraction of stable stars appears to drop 
precipitously near solar metallicity. Note, however, that the latter 
conclusion needs for an additional check since the most metal rich bin is 
only sampled by 6 stars.

\begin{figure}
\plotone{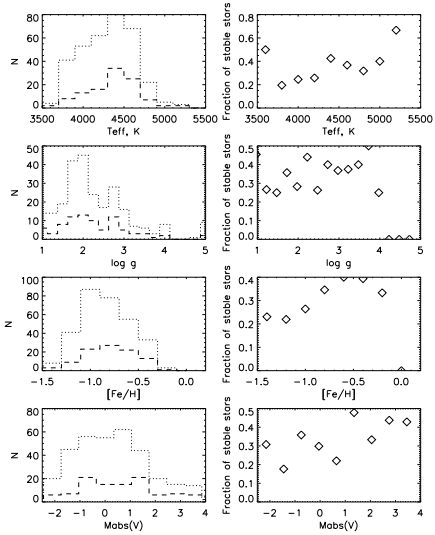}
\caption{
The same as in Figure \ref{fig9}, but for the southern subsample.
\label{fig9}
}
\end{figure}

An analog of Figure \ref{fig8} is drawn for the southern subsample
in Figure \ref{fig9}. 
Here we use the "photometric" [Fe/H] derived by \citet{rhee}. 
The surface gravity is established by way of interpolation of the
isochrones published by \citet{bergbusch92}.
The histograms for $T_{eff}$ defined by the same method for both
subsamples of northern and southern stars, are qualitatively similar in
Figures \ref{fig8} and \ref{fig9}. Though the other parameters are estimated
by different ways, the corresponding histograms for [Fe/H], $\log~g$, 
and $M_V$ are qualitatively rather similar. 

\begin{figure}
\epsscale{.8}
\plotone{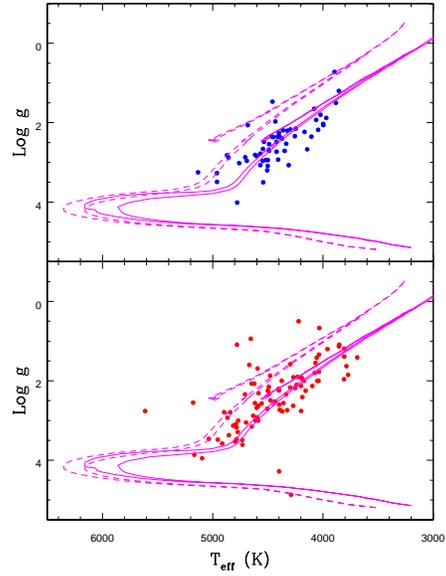}
\caption{
The spectroscopically derived surface gravity, log $g$, plotted versus 
T$_{\rm eff}$ for the McDonald northern sample: The top panel shows 
stars with $\sigma \,<$ 100 m s$^{-1}$ while the bottom panel shows 
those with RV-variability greater than 100 m s$^{-1}$. The RV-stable stars 
are more concentrated in the $\log~g$ -- $T_{eff}$ plane than the variables. 
The continuous curves are isochrones from \citet{girardi00}, with the dashed 
curves having [Fe/H]=-0.7 and the solid curves [Fe/H]=0.0.  The two 
isochrones for each metallicity have different ages of 3.5 Gyr and 
8.9 Gyr, respectively.  The concentrated ``clump'' of RV-stable stars may 
correspond to the core-He burning red giant phase of stellar evolution. 
\label{fig10}
}
\end{figure}

In order to better constrain what types of stars may be most RV-variable
we again focus on the McDonald data 
and the stellar parameters derived from these spectra.  In Figure \ref{fig10}
the values of $\log~g$ are plotted versus the effective temperatures: this
is a spectroscopic version of an HR-diagram.  The top panel contains
those stars with RV variability of less than 100 m s$^{-1}$, while the bottom
panel are the definite RV variables having $\sigma \,>$ 100 m s$^{-1}$.
Superimposed
on these stellar points are isochrone curves from \citet{girardi00};
two metallicity isochrones are shown (the solid curves have [Fe/H]=0.0
while the dashed curves have [Fe/H]=-0.7) with two ages for each
metallicity (3.5 and 8.9 Gyr).  
Plotted this way, the initial RV-results indicate that those stars 
exhibiting variability of less than $\sim$100 m s$^{-1}$ tend to be 
more concentrated in the $log~g \,-\, T_{eff}$ plane, near the He-core 
burning clump for near-solar metallicities. 
The stars tending to show RV-variability fall all along either 
well-defined first-ascent RGB, or AGB, both at somewhat low metallicities.  
Given uncertainties in both the models and the gravities and temperatures 
derived by us (which almost certainly also carry some systematic 
differences), the concentration of possibly RV-stable red giants 
near $T_{eff}$$\sim$4500K and $\log~g\sim$2.5-3.0 coincides 
quite closely to the red giant clump for stars having metallicities 
near solar, or slightly lower. Note that real $\log~g$ may have 
systematically lower values than those shown in Figure \ref{fig10}
(see also \S{3.4}). 

\begin{figure}
\epsscale{.8}
\plotone{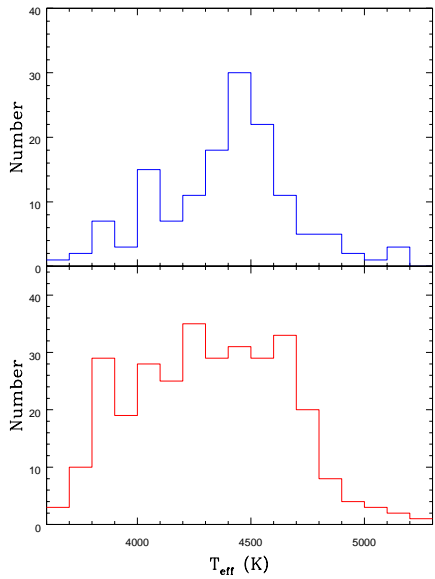}
\caption{
Effective temperature histograms segregated as the RV-stable stars 
(top panel) and RV-variable stars (bottom panel) for the combined
northern and southern data sets.  The RV-stable stars are sharply
peaked near $T_{eff}$=4500$\pm$200K, while the RV-variable
stars are more spread out in temperature.  This provides additional
support for the suggestion that the core-He burning stars, with only
mild metal deficiencies, may be intrinsically more stable in radial velocity. 
\label{fig11}
}
\end{figure}

The possible association of quasi-RV stability with the red giant clump can
be investigated further by including the southern sample, although we do not
have spectroscopically derived gravities for these stars.  For a relatively 
old population, 
however, the effective temperature of the clump does not
vary much if the metallicity does not vary by much: this can been seen by
inspecting the isochrones in Figure \ref{fig10}.  
The indications from the McDonald 
results are that the RV-stable red giants are only mildly metal poor with a
relatively small dispersion in metallicity.  
For the 36 McDonald RV-stable stars having
T$_{\rm eff}$=4500$\pm$200K (corresponding to the observed ``clump'' of
points in Figure \ref{fig10}), the mean metallicity 
is [Fe/H]=-0.5$\pm$0.2.  The combined northern and southern samples are
plotted as effective temperature histograms in Figure \ref{fig11}, 
where the sample
has been segregated using $\sigma$.  There is a clearly defined clump of
potentially RV-stable stars with T$_{\rm eff}$$\sim$4500K that probably
corresponds to core-He burning red giants with mild metal deficiencies.
The stars with $\sigma \,>$100 m s$^{-1}$, shown in the bottom panel,
are significantly more dispersed in T$_{\rm eff}$, with almost equal
numbers over a range in effective temperature; our suggestion is that
these are primarily first-ascent giants or AGB stars, as well as intrinsically
RV-stable stars (such as clump giants) that have companions.   
The fraction of stars whose RV-variability is caused by components is not 
known from our observations. Demonstrating the presence of companions will
require much more repeated observations.

Preliminary results from monitoring of photometric variability of a 
subsample of GGSS stars by \citet{biz04} reveal no correlation between RV- 
and brightness variations. However, more precise and {\it simultaneous} 
photometric and spectroscopic observations are needed to shed light on 
the nature of the RV variations in red giants.

\subsection{Strategies for Selecting SIM Astrometric Grid Stars}

Our analysis gives insight into the optimal selection of giant stars for the 
SIM Astrometric Grid if time and resources are limited (as indeed they are) 
for ground-based vetting of large numbers of stars for the most RV-stable 
specimens.

As can be seen in Figure \ref{fig11},
a higher fraction of stars with $\sigma \,<\, 100$ m s$^{-1}$ can be found
with $ 4300\,K \,<\, T_{eff} \,<\, 4700\,K$.  This is a simple, 
straightforward
selection criterion resulting in an initial success rate of 50\% in finding
RV-stable red giants.  As a point of reference, the entire northern sample
contains 39\% stable stars and the southern sample contains 32\%.  If
high-resolution spectra are also used to derive surface gravities and
metallicities, the additional constraints of  
-0.5 $<$ [Fe/H] $<$ -0.1, and $2.3 ~<~  log~g ~<~ 3.2$ lead to a 81\% 
and 40\% success rate for the northern and southern samples, respectively, 
yet significantly narrows the sample.
Note that the northern sample restricted simultaneously by all stellar 
parameters reveals the fraction of RV-stable stars that is close to 
findings by \citet{frink01}. 


More attention to exploring the solar and higher metallicity ranges is 
warranted because if the relatively low RV-stable fraction we find at these
solar-like metallicities is borne out it would have serious ramifications
for the currently selected SIM Astrometric Grid sample, which is dominated
by giant stars randomly selected from the Tycho catalogue. Spectroscopic
metallicities measured for nearly 500 Tycho giants by J. Crane (in
preparation) in the preferred magnitude range ($V<11$) for the SIM Grid
stars indicates a median metallicity for this sample of [Fe/H] = -0.1,
and perhaps larger (i.e., 0.0 dex) if the arguments for the need of
a +0.1 dex correction to the local metallicity scale are
applied \citep{reid02, haywood02}. Thus, a significant fraction of the
Tycho sample is at metallicities where a large fraction of stars 
may have troublesome atmospheric jitter.
   
\section{Conclusions}

We are carrying out high-resolution spectroscopic observations of an all-sky
sample of giants which was pre-selected based on the GGSS program of
Washington photometry. The GGSS was intended to identify stars with the
highest likelihood of astrometric stability --- namely, subsolar metallicity
red giants.  The main purpose of the present study is to make an initial 
assessment of the
RV-variability properties of GGSS stars as a function of their atmospheric
characteristics. The RV variation was estimated for 489 stars with
spectroscopy taken on both Northern and Southern telescopes. Basic stellar
parameters were estimated for the northern subsample from high-resolution
spectroscopy. 

A surprisingly high fraction of investigated giant stars have unstable 
radial velocities at 100 m s$^{-1}$ level: about 2/3 of our sample. 
Both of the samples, northern and southern, although having been
observed independently and with different instruments and techniques,
show similar distributions of RV variability.  Although a number of
obvious spectroscopic binaries are included in this sample, much of
the low-amplitude RV variability is probably due to atmospheric
motions and temperature inhomogeneities.  

A higher fraction of stars with $\sigma < 100$ m s$^{-1}$ can be found 
among the objects with $4300\,K < T_{eff} < 4700\,K$, which corresponds 
to $(J-K)_0$ = 0.59 -- 0.73. If we incorporate spectroscopic metallicity 
and surface gravity information as well, and select only stars with 
-0.5 $<$ [Fe/H] $<$ -0.1 and $2.3 <  \log~g < 3.2$, it will help to identify 
RV-stable candidates more effectively, but narrows the sample of potential 
candidates substantially. From the point of view of minimizing expensive
observations, the optimal way of preliminarily selecting RV-stable stars is 
to use the effective temperature, or more exactly, the calibration independent 
range of NIR colors as $0.59 < (J-K)_0 < 0.73$.  This range encompasses 44\%
of stars with $\sigma < 100 \, m\, s^{-1}$ in both our southern and northern 
samples; 38.5\% of the initial GGSS sample of candidate Astrometric Grid stars
(1710 of 4440) lies in this range. The result obtained here will aid 
in the continuing efforts to define the astrometric reference grid for SIM.

\begin{acknowledgments}

The GGSS follow-up observations presented here have been
supported by the JPL and NASA via grant 99-04-OSS-058.
The photometric portion of the GGSS was funded by NASA/JPL grants 1201670 
and 1222563 to SRM and RJP, and with substantial contributions of additional 
funding from a David and Lucile Packard Foundation Fellowship to SRM.
VVS acknowledges support for part of this work
from NASA via grant NAG5-13175.
We are grateful to the staff of McDonald observatory for their help and
support of both photometric and high resolution spectroscopic observations, and from the
Carnegie Observatories for similarly generous contributions for photometric and medium resolution
spectroscopic observations of the GGSS. 
This publication makes use of data products from the Two Micron All Sky
Survey, which is a joint project of the University of Massachusetts and the
IPAC/Caltech, funded by the NASA and NSF.

\end{acknowledgments}

\clearpage


\begin{deluxetable}{lcccccccccc}
\tabletypesize{\scriptsize}
\tablewidth{0pt}    
\tablecaption{Stellar Parameters for the entire McDonald Sample of 434 GGSS stars.}

\tablehead{ \colhead{Name} & \colhead{RA~(J2000.0)} & \colhead{DEC~(J2000.0)} &
\colhead{$T_{eff}$} & \colhead{[Fe/H]$_{ph}$} & 
\colhead{D$_{ph}$} & \colhead{$\log~g$$_{sp}$} & 
\colhead{$\Delta \log~g$$_{sp}$} & \colhead{[Fe/H]$_{sp}$} & 
\colhead{$\Delta$[Fe/H]$_{sp}$} \\
\colhead{} & \colhead{(HH:MM:SS)} & \colhead{(DD:MM:SS)} &
\colhead{K} & \colhead{(dex)} & \colhead{(Kpc)} & \colhead{(dex)} &
\colhead{(dex)} &\colhead{(dex)}&\colhead{(dex)}
}
\startdata
G2358+00.31 & 00:00:51.96 & 00:22:16.8 & 4290 & -0.3 & 1.3 & 4.87 & 0.19 & -2.80 & 0.05\\
G2358+00.92 & 00:01:36.38 & 00:14:09.9 & 4399 & -0.9 & 1.1 & 2.24 & 0.13 & -0.42 & 0.05\\
G0001+00.94 & 00:04:20.08 & 00:27:12.2 & 4324 & -0.7 & 3.0 & 2.67 & 0.41 & -0.40 & 0.16\\
G0011+05.87 & 00:14:18.85 & 05:57:37.5 & 4440 & -0.5 & 0.7 & 2.13 & 0.19 & -0.47 & 0.09\\
G0011+16.75 & 00:14:26.65 & 17:16:03.0 & 4594 & -0.8 & 1.8 & 1.47 & 0.38 & -0.67 & 0.19\\
...&&&&&&&&&\\
\enddata
\tablecomments{Name of star, its RA and DEC,
effective temperature, photometric metallicity [Fe/H], photometric distance,
spectroscopic surface gravity $\log~g$ and its accuracy, and spectroscopic 
metallicity and its accuracy.\\
{\bf The full version of the Table is presented in electronic form.}
}
\label{tab1}
\end{deluxetable}


\begin{deluxetable}{ ccccc }  
\tablewidth{0pt}  
\tablecaption{The \ion{Fe}{1} \& \ion{Fe}{2} Lines}

\tablehead{  $\lambda$($\AA$) &   
\multicolumn{1}{c} {Species} &
\multicolumn{1}{c} {$\chi$(eV)} &
\multicolumn{1}{c} {$\log{gf}$} 
} 
\startdata 
5307.361  & Fe I & 1.608 &   -2.970 \\                     
5322.041  & Fe I & 2.279 &   -2.840 \\                       
5497.516  & Fe I & 1.011 &   -2.840 \\                     
5501.464  & Fe I & 0.958 &   -2.957 \\                   
5522.447  & Fe I & 4.209 &   -1.400 \\                     
5536.583  & Fe I & 2.832 &   -3.812 \\                        
5539.284  & Fe I & 3.642 &   -2.660 \\                         
5549.948  & Fe I & 3.695 &   -2.904 \\                         
5559.638  & Fe I & 4.988 &   -1.829 \\                       
5560.207  & Fe I & 4.435 &   -1.040 \\                       
5577.031  & Fe I & 5.033 &   -1.551 \\                          
5579.335  & Fe I & 4.231 &   -2.406 \\                      
5607.664  & Fe I & 4.154 &   -2.258 \\                       
5608.974  & Fe I & 4.209 &   -2.402 \\                       
5611.361  & Fe I & 3.635 &   -2.993 \\ 
5618.631  & Fe I & 4.209 &   -1.260 \\
5619.224  & Fe I & 3.695 &   -3.256 \\ 
5636.696  & Fe I & 3.640 &   -2.608 \\
5638.262  & Fe I & 4.220 &   -0.720 \\ 
5661.012  & Fe I & 4.580 &   -2.432 \\ 
5677.684  & Fe I & 4.103 &   -2.694 \\ 
5679.025  & Fe I & 4.652 &   -0.770 \\ 
5696.103  & Fe I & 4.549 &   -1.997 \\ 
5698.023  & Fe I & 3.640 &   -2.689 \\                       
5705.466  & Fe I & 4.301 &   -1.360 \\                       
5717.835  & Fe I & 4.284 &   -0.980 \\ 
5724.454  & Fe I & 4.284 &   -2.627 \\                       
5759.261  & Fe I & 4.652 &   -2.073 \\                       
5760.345  & Fe I & 3.642 &   -2.490 \\                       
5784.657  & Fe I & 3.397 &   -2.673 \\                       
5793.913  & Fe I & 4.220 &   -1.697 \\                       
5806.717  & Fe I & 4.608 &   -0.900 \\                       
5807.782  & Fe I & 3.292 &   -3.404 \\                       
5809.217  & Fe I & 3.884 &   -1.690 \\                       
5811.917  & Fe I & 4.143 &   -2.427 \\                       
5814.805  & Fe I & 4.283 &   -1.820 \\                       
5837.700  & Fe I & 4.294 &   -2.337 \\                       
5838.370  & Fe I & 3.943 &   -2.337 \\                      
5844.917  & Fe I & 4.154 &   -2.940 \\                      
5845.266  & Fe I & 5.033 &   -1.818 \\                      
5849.682  & Fe I & 3.695 &   -2.993 \\                       
5853.149  & Fe I & 1.485 &   -5.268 \\                     
5856.083  & Fe I & 4.294 &   -1.640 \\ 
5861.107  & Fe I & 4.283 &   -2.452 \\                       
5000.743  & Fe II & 2.778 &   -4.745 \\                      
5018.440  & Fe II & 2.891 &   -1.213 \\                       
5132.669  & Fe II & 2.807 &   -4.000 \\                       
5234.625  & Fe II & 3.221 &   -2.240 \\                       
5264.812  & Fe II & 3.230 &   -3.188 \\                       
5284.098  & Fe II & 2.891 &   -3.010 \\                       
5325.559  & Fe II & 3.221 &   -3.170 \\                       
5414.046  & Fe II & 3.221 &   -3.620 \\                       
5425.247  & Fe II & 3.199 &   -3.210 \\                       
5534.847  & Fe II & 3.244 &   -2.770 \\                       
\enddata
\tablecomments{Adopted excitation potentials and gf-values 
for \ion{Fe}{1} and \ion{Fe}{2} lines}
\label{tab2}
\end{deluxetable}


\begin{deluxetable}{ cccccccccccc }  
\tablewidth{400pt}  
\tablecaption{Equivalent-Width Measurements}

\tablehead{  $\lambda$ &   
\multicolumn{1}{c} {\#1} &
\multicolumn{1}{c} {\#2} &
\multicolumn{1}{c} {\#3} &
\multicolumn{1}{c} {\#4} &
\multicolumn{1}{c} {\#5} &
\multicolumn{1}{c} {\#6} &
\multicolumn{1}{c} {\#7} &
\multicolumn{1}{c} {\#8} &
\multicolumn{1}{c} {\#9} &
\multicolumn{1}{c} {\#10} 
} 
\startdata 
5307.361  &  132 & 109 & 185 & 117 & 118 & 154 & 154 & 165 & 116 & 135 \\                     
5322.041  &   92 &  84 & 140 & 86  & 85  & 114 & 100 & 118 & 819 & 102 \\                      
5497.516  &  ... & 183 & ... & 172 & ... & ... & ... & ... & ... & ... \\                     
5501.464  &  ... & ... & ... & 166 & ... & ... & ... &  ... &  ... &  ... \\                   
5522.447  &  ... &  50 & ... &  47 &  ... & 73 &  ... &  ... &  ... &  ... \\                     
5536.583  &   35 &  25 & ... &  17 &  ... & 34 &  ... &  ... &  ... &  ... \\                        
5539.284  &   42 & ... & 54  &  23  & 32   & 45 & 45   & 44   & 28   &  ... \\                         
5549.948  &   32 &  15 & 44  &  ... &  ... &  ... &  ... & 31 & 21 &  ...  \\                        
5559.638  &   18 & ... & 19  &  ... &  ... &  ... & 15 & 11 &  ... &  ...  \\                     
5560.207  &   63 & ... & ... &  57  &   60 & 72 & 68 & 68 & 56 & 62   \\                    
5577.031  &  ... & ... & 22  &  ... &  ... &  ... &  ... &  ... & 15 &  ... \\                         
5579.335  &  ... &  14 & 34  &  10  &  ... &  ...  & 25   & 23   & 16 &  ... \\                       
5607.664  &  ... & ... & 44  &  ... & 25   &  ... & 35 & 33 & ... & 27 \\                      
5608.974  &   25 & ... & 35  &  ... & ... & 24 & 31 & ... &  ... & 13  \\                     
5611.361  &   31 & ... & 37  &  14  & 19  & ... &  37 & 27 &  ...&  19 \\
5618.631  &   70 &  58 & 91  &  56  & 62  & 78 & 71 & 76 & 58 & 64 \\
5619.224  &   21 &  11 & 195 &  6   & 10  & 15 & 21 & 15 & ... &  ... \\
5636.696  &   40 & ... & 65  &  24  & ... & 48 &  ... & 40 & 32 &  ... \\
5638.262  &   98 &  78 & 126 &  78  & 89  & 99 &  92 &  98 &  77 &  96 \\
5661.012  &  ... &  ... &  15 &  ... &  ... &  ... & 13 &  ... &  ... &  ... \\
5677.684  &   20 &  ...  & 24 &  ... & 15 & 17 & 24 & 20 &  ... & ... \\
5679.025  &   66 &  54  &  ... & 53 & 63 & 74 & 72 & 74 & 51 & 66 \\
5696.103  &   27 &  17  &  ... &  ... &  ... &  22  &  ...  &  ...  & 18  & 18 \\
5698.023  &  ... &  28  & 50 &  ... & 31  & 39  & 47   & ...  &  ...  & 30      \\                  
5705.466  &   50 &  48  & 88 & 44 & 54 &  63 &  65 &   ... &  47 &  52   \\               
5717.835  &   76 &  69 & 105  & 66 &  67 &  86 &  81 &  82  &  ... &  72 \\                      
5724.454  &   19 &  ... &  19 &   ... &   ... &   ... &   ... &   ... &  14 &   9   \\                     
5759.261  &  ... &  ...  & 20  &  ...  & 13  &  ...  & 50  & 15  & 12  & 11    \\                    
5760.345  &   44 &  ... &  65 &  29 &  41 &  53 &  23 &   ... &  38 &  36 \\                     
5784.657  &   52 &  36  & 77 & 34 & 42 & 54 &  57 &  60 &  44 &  41 \\                      
5793.913  &   51 &  38  & 71 & 39 & 46 & 55 & 54 &  57 &  42  &  ...   \\                     
5806.717  &   60 &  55  & 90 & 49 & 56 & 68  & 66  & 66  &  ...  & 64  \\                      
5807.782  &   27 &  ... &  41 & 16 & 23 & 26 &  39 &  30 &  17 &   ...   \\                     
5809.217  &   70 &  54  & 98 & 53 & 56 & 76 &  70 &  79  &  ... &  70   \\                     
5811.917  &  ... &  15  & 35 & ... &  ... &  ... &  29 &  25 &  16 &   ...      \\                  
5814.805  &   36 &  34  & 59 &  ... & 37 &  ... &  48 &  42 &   ... &  33    \\                    
5837.700  &  ... &  13  & ... & 12 & 16 &  ... &  28 &  21  &  ... &   ...   \\                     
5838.370  &  ... &  25  &  ... & 21 & 32 &  ... &  45  &  ... &   ... &  31   \\                    
5844.917  &  ... &  ... &  ... &  ... &  ... &  ... &  16 &   ... &   8 &   8   \\                    
5845.266  &  ... &  ... & ...  & ... &  ... & 10 &  14 &   ... &   9 &   10    \\                   
5849.682  &  ... &  14  & 31  & ... & 17 & 20 &  30 &  26 &  19 &  14    \\                    
5853.149  &  ... &  32   & 67 & 22 & 32 & 55 &  58 &  57  &  ... &  27   \\                     
5856.083  &   51 &  ...  & 70 &  ... & 42 & 51 &  53 &  56 &  39 &   ...   \\                     
5861.107  &  ... &  ...   & ... &  ... &  ... &  ... &  22.7 &   ... &   ... &   ...   \\                     
5000.743  &  ... &  ...   & ... &  ... &  ...  & ... &   ... &   ...  &  ... &   ...   \\                    
5018.440  &  ... & 175  & 201  & ... &  ...& 222 &  ... & 186  & 174 & 161    \\                    
5132.669  &   39 &  ...  & 37 & 41 & 35 & 49 &  45 &   ... &   ... &  25   \\                     
5234.625  &   87 &  71  & 100 & 92 & 72 & 100 &  86 &  81 &  72 &  786   \\                     
5264.812  &   45 &  43  & 55 & 54 & 44 &  57 &  56 &  44 &  38 &  42    \\                    
5284.098  &   64 &  63   & 82 & 76 &  ...  &  ... &   ... &  67 &  62 &   ...   \\                     
5325.559  &  ... &  42  & 54 & 56 & 47  & 57 &  56 &  45 &   ... &  36   \\                     
5414.046  &   30 &  30  & 30 & 37 & 34  & ... & 42 &  30 &  29 &  20   \\                     
5425.247  &   46 &  41  & 52 & 57 & 46 & 61 & 51 & 44 & 38 &  38   \\                     
5534.847  &  ... &  60  & 70 & 73 & 59 & 74 & 70  &  ...  &  ...  &  ...   \\                     
\enddata
\tablecomments{Equivalent widths of \ion{Fe}{1} and \ion{Fe}{2} lines 
(given in m\AA) for 10 stars. Stellar identifications are provided in 
Table 4.
}
\label{tab3}
\end{deluxetable}


\begin{deluxetable}{ ccccccc }  
\tablewidth{0pt}  
\tablecaption{Stellar Parameters and Metallicities derived from classical spectroscopic analysis for 10 selected stars}

\tablehead{  Star &   
\multicolumn{1}{c} {ID} &
\multicolumn{1}{c} {T$_{\rm eff}$} &
\multicolumn{1}{c} {$log~g$} &
\multicolumn{1}{c} {$\xi$} &
\multicolumn{1}{c} {[Fe/H]} \\
 & & (K) & (dex) & (km/s) & (dex)
} 
\startdata 
G0858+00.192 & \#1 & 4630 & 2.2 & 1.1 & -0.31 $\pm$0.07 \\        
G0901+00.175 & \#2 & 4760 & 2.6 & 0.9 & -0.48 $\pm$0.06 \\ 
G1007+00.5   & \#3 & 4490 & 2.5 & 1.8 & -0.12 $\pm$0.05 \\ 
G1018-05.46  & \#4 & 4670 & 1.6 & 1.2 & -0.77 $\pm$0.05 \\
G1030+00.15  & \#5 & 4780 & 2.6 & 0.9 & -0.34 $\pm$0.06 \\
G1042-05.69  & \#6 & 4370 & 1.1 & 1.4 & -0.63 $\pm$0.06 \\
G1053+00.4   & \#7 & 4400 & 1.2 & 0.9 & -0.35 $\pm$0.07 \\ 
G1113+00.20  & \#8 & 4240 & 1.6 & 1.4 & -0.56 $\pm$0.06 \\
G1147-05.31  & \#9 & 4600 & 2.1 & 0.6 & -0.48 $\pm$0.07 \\
G1147-05.49  & \#10 & 4720 & 3.0 & 1.3 & -0.41 $\pm$0.08 \\
\enddata
\tablecomments{Name of star, its ID in Table 3, 
effective temperature, surface gravity, microturbulent velocity, and
[Fe/H].
}
\label{tab4}
\end{deluxetable}



\begin{deluxetable}{lcccccccccc}
\tabletypesize{\scriptsize}
\tablewidth{0pt}    
\tablecaption{RV-variability for the 148 McDonald Repeated Stars}
\tablehead{ \colhead{Name} & \colhead{M} & \colhead{$T_{\rm eff}$} &
\colhead{D$_{ph}$} & \colhead{[Fe/H]$_{ph}$}  &\colhead{$\log~g_{sp}$}& 
\colhead{$\Delta \log~g_{sp}$} & \colhead{[Fe/H]$_{sp}$} & 
\colhead{$\Delta$ [Fe/H]$_{sp}$} & \colhead{$\sigma$} & \colhead{N}\\
\colhead{} & \colhead{(mag)} & \colhead{(K)} & \colhead{(Kpc)} & \colhead{(dex)} &
\colhead{(dex)} & \colhead{(dex)} &\colhead{(dex)} &\colhead{(dex)} &\colhead{(km/s)} &
\colhead{}
}
\startdata
G0118-05.38  & 11.97 & 4814 & 0.9 & -1.0 & 3.13 & 0.13 &  0.00 & 0.05 & 0.165 &2\\
G0131+00.86  & 11.55 & 4707 & 0.9 & -1.2 & 2.34 & 0.17 & -0.80 & 0.14 & 0.919 &2\\
G0142+05.56  & 10.91 & 3959 & 2.2 & -0.9 & 1.20 & 0.65 & -0.60 & 0.28 & 0.489 &2\\
G0142+05.70  & 12.27 & 4196 & 2.3 & -0.6 & 1.93 & 0.25 & -0.67 & 0.25 & 0.287 &2\\
G0151+00.72  & 11.29 & 4334 & 1.7 & -0.6 & 2.34 & 0.06 & -0.07 & 0.22 & 0.287 &3\\
...&&&&&&&&&&\\
\enddata
\tablecomments{Name of star, Washington $M$ magnitude (which is 
approximately Johnson's $V$), effective temperature, photometric distance, 
photometric metallicity [Fe/H], spectroscopic surface gravity $\log~g$ and 
its accuracy, spectroscopic metallicity [Fe/H] and its accuracy, RV 
variability, and number of RV-observations for the northern sample objects.\\
{\bf The full version of the Table is presented in electronic form.}
}
\label{tab5}
\end{deluxetable}


\begin{deluxetable}{lcccccc}
\tablewidth{0pt}
\tablecaption{RV-variability for the 341 CORALIE Repeat Observation Stars}

\tablehead{ \colhead{Name} &\colhead{M} &\colhead{$T_{eff}$} & 
\colhead{D$_{ph}$} & \colhead{[Fe/H]$_{ph}$} & 
\colhead{$\sigma$} & \colhead{N} \\
\colhead{} & \colhead{(mag)} & \colhead{(K)} & \colhead{(Kpc)} & 
\colhead{(dex)} & \colhead{(km/s)} & \colhead{}
}
\startdata
G0000-56.81   &  11.92 & 4066 &  3.5 & -1.1 &  0.004 &  3 \\
G0012-28.38   &  11.71 & 3891 &  3.8 & -1.0 &  0.182 &  3 \\
G0016-39.1207 &  11.46 & 4418 &  4.3 & -2.0 &  1.104 &  7 \\
G0016-39.2075 &  11.93 & 4580 &  1.7 & -1.0 &  0.635 &  2 \\
G0016-39.3290 &  11.41 & 4753 &  0.8 & -0.7 &  2.060 &  4 \\
...&&&&&&\\
\enddata
\tablecomments{Name of star, Washington $M$ magnitude, effective temperature, 
photometric distance, photometric metallicity [Fe/H], RV variability, 
and number of RV observations for objects of the southern sample.\\
{\bf The full version of the Table is presented in electronic form.}
}
\label{tab6}
\end{deluxetable}

\clearpage

\end{document}